# Interactive and Automatic Generation of Primitive Custom Circuit Layout Using LLMs

Geunyoung You
Hanyang University
dbrmsdud22@hanyang.ac.kr

Youjin Byun
Hanyang University
yoojin234@hanyang.ac.kr

Sojin Lim
Hanyang University
gch30725@hanyang.ac.kr

Jaeduk Han
Hanyang University
jdhan@hanyang.ac.kr

***Abstract*—In this study, we investigate the use of Large Language Models (LLMs) for the interactive and automated production of customs circuit layouts described in natural language. Our proposed layout automation process leverages a template-and-grid-based layout generation framework to create process-portable layout generators tailored for various custom circuits, including standard cells and high-speed mixed-signal circuits. However, rather than directly describing the layout generators in traditional programming language, we utilize natural language using LLMs to make the layout generation process more intuitive and efficient. This approach also supports interactive modifications of the layout generator code, enhancing customization capabilities. We demonstrate the effectiveness of our LLM-based layout generation method across several custom circuit examples, such as logic standard cells, a serializer and a strong arm latch, including their completeness in terms of Design Rule Check (DRC), Layout Versus Schematic (LVS) test, and post-layout performance for high-speed circuits. Our experimental results indicate that LLMs can generate a diverse range of circuit layouts with substantial customization options.**

## I. Introduction

### *A. Trends in Circuit Design*

In recent years, advances in semiconductor technology have resulted in an exponential increase in the scale of Integrated Circuits (IC), posing significant challenges to the scalability and reliability of circuit design flows due to increased manufacturing and design costs. These challenges adversely affect the performance of custom circuits, including Analog and Mixed-Signal (AMS) circuits as well as primitive cells for custom digital and memory circuits, due to their high sensitivity to the device characteristics and post-layout effects. Consequently, numerous developments of Electronic Design Automation (EDA) are underway incorporating advanced algorithms and technologies.

In particular, new approaches to design and produce integrated custom circuit layouts with enhanced productivity are gaining prominence. Various Python-based layout generation engines have been developed to facilitate the creation of custom circuit layouts in advanced CMOS manufacturing processes. They respond to the increasing demand for high design productivity in physical layouts by providing frameworks for writing layout generators in Python [1]-[4]. However, the generator-based layout generation frameworks often require significant time and are susceptible to human errors, as engineers must describe tasks in programming language. Therefore, we aim to decrease errors and enhance productivity and customization capability by enabling engineers to represent their intentions in natural language and produce actual generator code using LLMs.

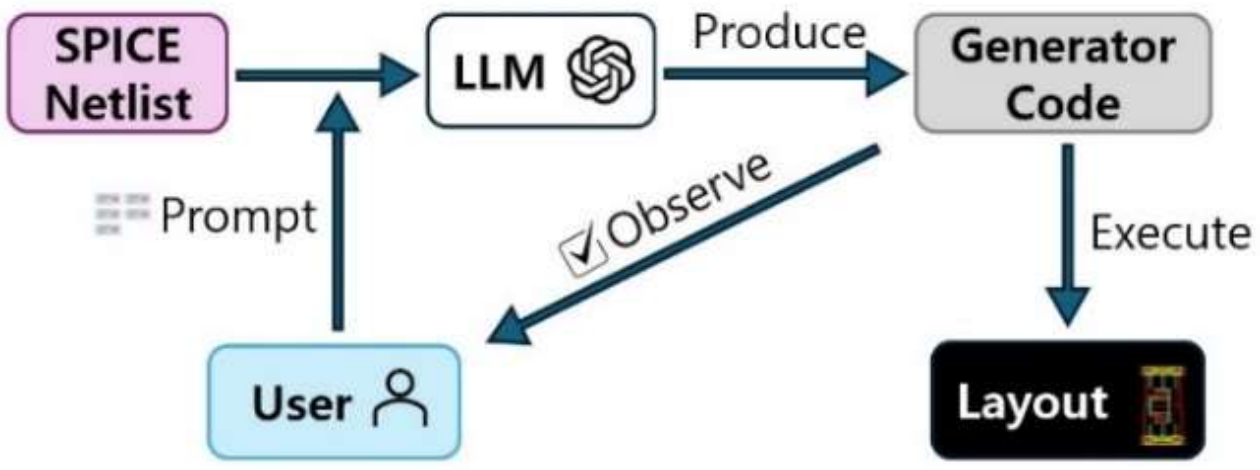

Fig. 1. Applying an interactive LLM to layout design.

### *B. Layout Generation Engine*

The template-and-grid-based layout generation [2], which we utilized as the baseline methodology, support rapid generation by placing customized templates on process-specific placement grids. It is specifically optimized for layout generation in nanometer technologies using various techniques, including template-and-grid-based generation, relative placement, and advanced array manipulation capabilities. This method provides designers with a rich, user-friendly experience in custom layout creation. The layout generation framework was developed in the Python programming language, which offers superior code readability and high productivity, enabling designers to easily develop and update layout generators if they are familiar with the Python programming language.

However, modern circuit designers employing the template-and-grid-based layout generation framework must undertake the process of translating the user's desired operations into code for each layout generator. The layout generators' code, based on Python, primarily focuses on creating layout objects on abstract coordinates for process portability. This approach makes it difficult for designers to quickly comprehend the content. Furthermore, it is not intuitively structured to reflect the intentions of circuit designers, who may not be proficient in Python. These complexities add significant barriers to the widespread adoption of layout automation.

Therefore, if the layout generation engine understands desired generation/editing tasks described in natural language, they can be executed more intuitively and effectively. A conversion engine, such as an LLM, can facilitate this by interpreting the designer's intentions expressed in natural language, converting them into executable code compatible with the aforementioned software framework. Additionally, if the LLM contributes to the preliminary optimization process for the desired electrical and physical characteristics, this could herald a revolutionary approach to design optimization as well as automation. Leveraging Artificial Intelligence (AI) in this way represents a new paradigm in circuit design, potentially reducing significant design efforts and costs for physical layout design (Fig. 1).

### C. *Contributions of Proposed Work*

Our contributions are summarized as follows:

• Presenting early examples of creating layout of custom circuits with an interactive LLM.

• Conducting a study on the entire process of designing a layout for a custom circuit using ChatGPT-4 [6].

• Providing practical guidelines for the effective application of interactive LLMs to circuit layout-related tasks.

• Deriving basic layout design optimization using AI, achieve important measures for circuit design.

## II. Background and Related Work

### A. *Large Language Models (LLMs)*

With the advent of AI, especially through the development of LLMs like ChatGPT from OpenAI and BARD from Google, research in human-computer interactions has significantly advanced [7], [8]. These models have demonstrated remarkable abilities to generate accurate responses across various domains. LLMs are typically constructed using variants of the transformer architecture, which employs self-attention mechanisms to efficiently handle large volumes of sequential data. The capabilities of LLMs have propelled forward advancements in automated content creation, conversational systems, and complex data interpretation tasks.

### B. *LLMs for Circuit Design*

Machine Learning (ML) has been extensively applied to circuit design, encompassing areas such as design space exploration, logic optimization, macro placement/floor planning, placement and routing, clock tree synthesis, and physical verification. Particularly, Deep Learning (DL) has played a crucial role in the development of Electronic Design Automation (EDA) tools throughout the chip design flow [9]. However, LLMs have not been widely utilized for chip design, with most previous applications focusing on generating Hardware Description Language (HDL) code. For instance, Pearce et al. [10] fine-tuned the GPT-2 model using synthetically generated Verilog HDL code and assessed the model on 'undergraduate-level' tasks. Nevertheless, LLMs have seldom been applied directly to highly specialized tasks such as circuit layout design. To efficiently utilize LLMs for a specific objective, a code-based framework relevant to that objective must be established in advance, since LLMs are highly effective in generating software code. Therefore, for layout generation, we have employed the template-and-grid based layout generation framework [2] that allows for the creation of layouts interactively and/or automatically with the aid of LLMs.

### C. *Layout Generation Techniques*

Over the past few decades, various techniques have been explored for AMS and custom circuit layout generation. Some of such approaches involve combining placement and routing algorithms to facilitate analog layout synthesis. For example, ALSYN [11] offers user-defined rule descriptions that guide algorithms in grouping, placing, and routing objects. Additionally, there are template-based software approaches ([12] and [13]) that generate layouts for analog functional blocks by merging generator codes with user-provided technical specifications. While these methods typically assume relatively simple and consistent design rules, there are also layout generation approaches designed for advanced technology nodes [14].

## III. LLM-Assisted Layout Generation

To effectively capture the subtle and complex requirements on physical design of custom circuits and translate them into practical layouts, we utilized LLMs to produce Python code compatible with the aforementioned template-and-grid-based layout generation framework. Generating Python code is one of the tasks that LLMs can perform very effectively, and they are already used by over 10,000 software developers for code generation, demonstrating a positive impact on its productivity [15].

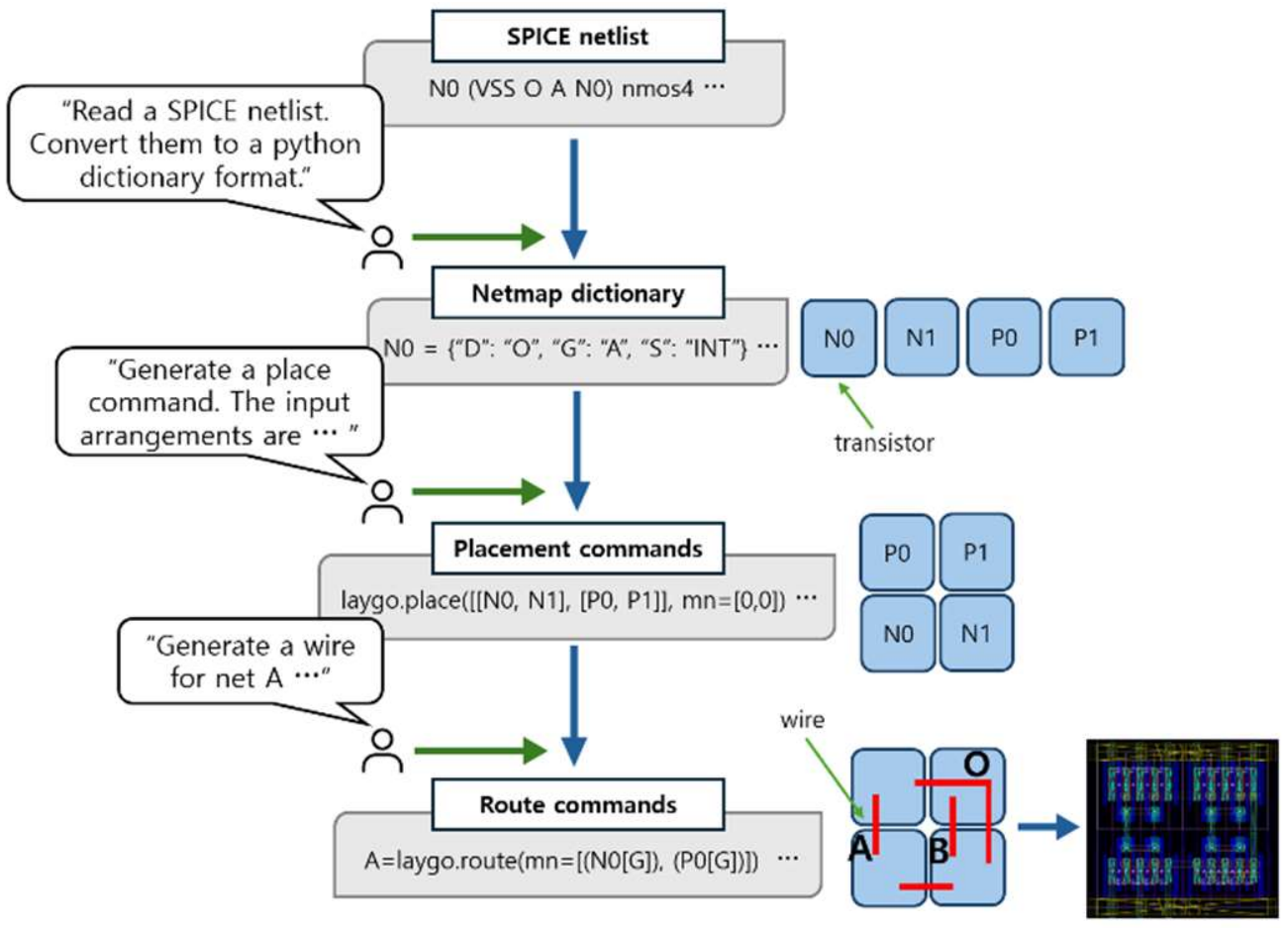

Fig. 2. Proposed LLM-based layout generation process.

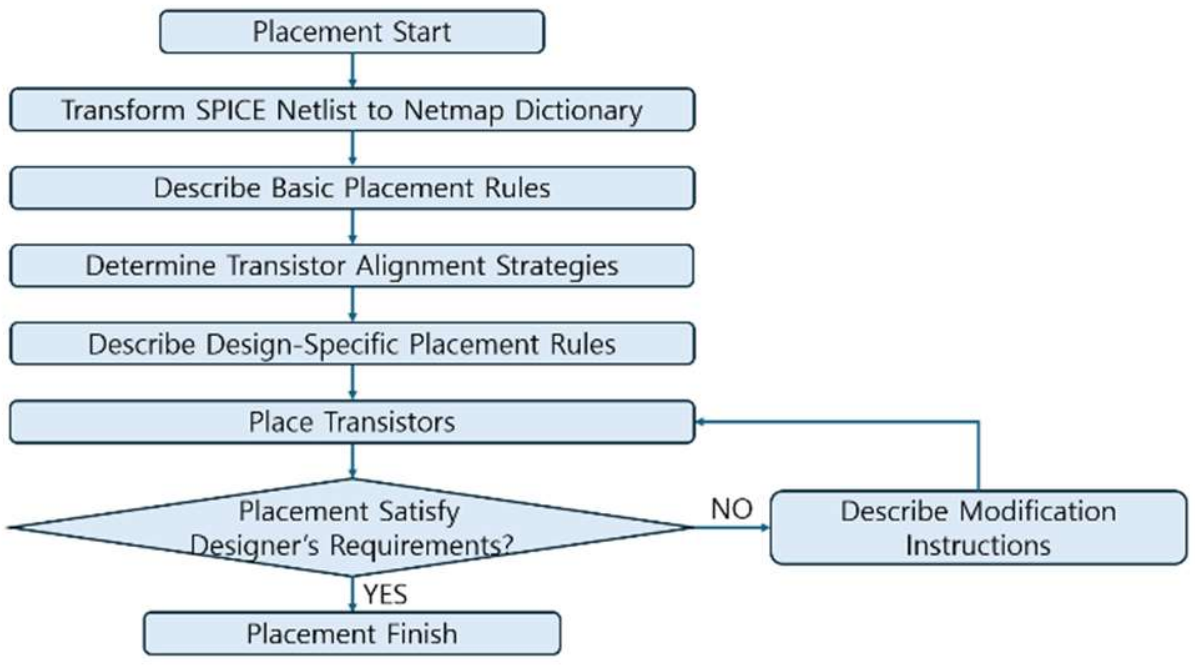

Fig. 3. Transistor-level placement.

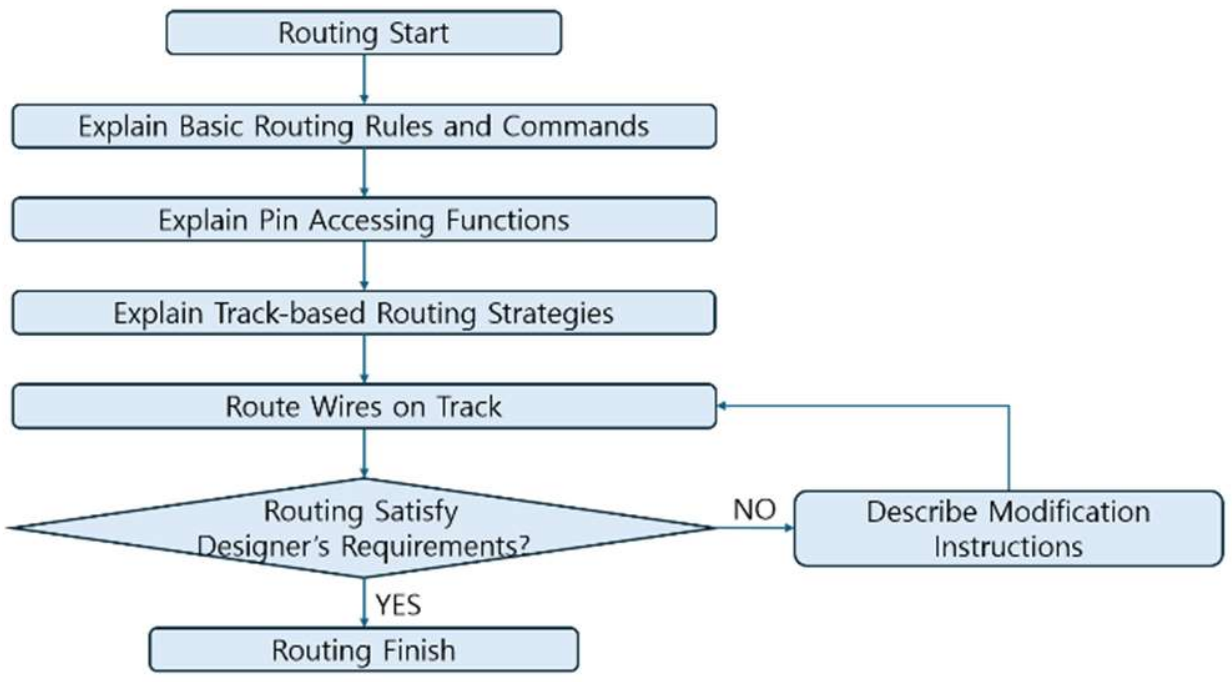

Fig. 4. Transistor-level routing.

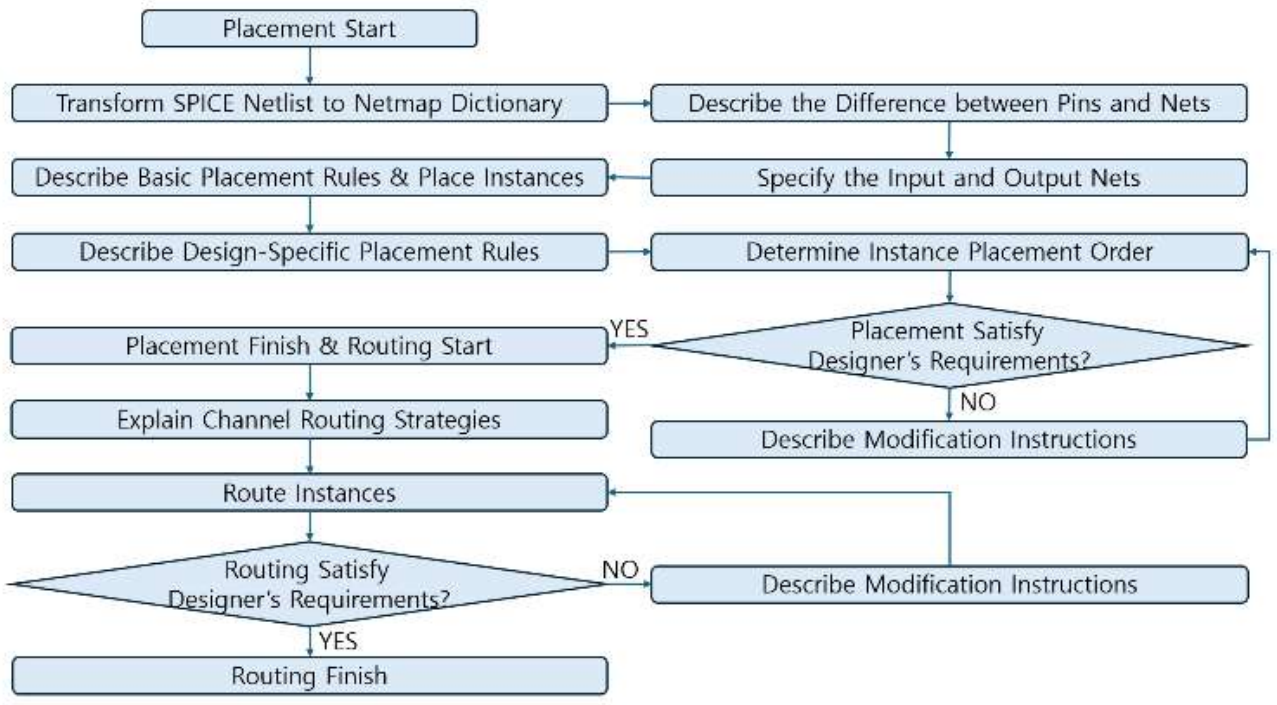

Fig. 5. Gate-level placement and routing.

In particular, our approach interactively supports layout generation, making it easy to modify or optimize characteristics by selectively incorporating design-specific conditions and generating or modifying associated Python statements. We aim to promote generative AI-based methods in the layout design domain, such that engineers produce layouts and qualitatively assess the results using competent commercial LLMs (such as ChatGPT-4 in OpenAI), as presented in Fig. 2.

The process initiates with an LLM analyzing the input SPICE netlist, including its device and connectivity information. Users then specify placement and routing instructions using natural language, and optionally add constraints and objectives to optimize characteristics such as device placement order and total wire length. To systematically produce and enhance the layout, we proceed based on the algorithms for various types of tasks, which are depicted in Fig. 3-5. Based on these algorithms, the target layout can be generated with minimal prompts.

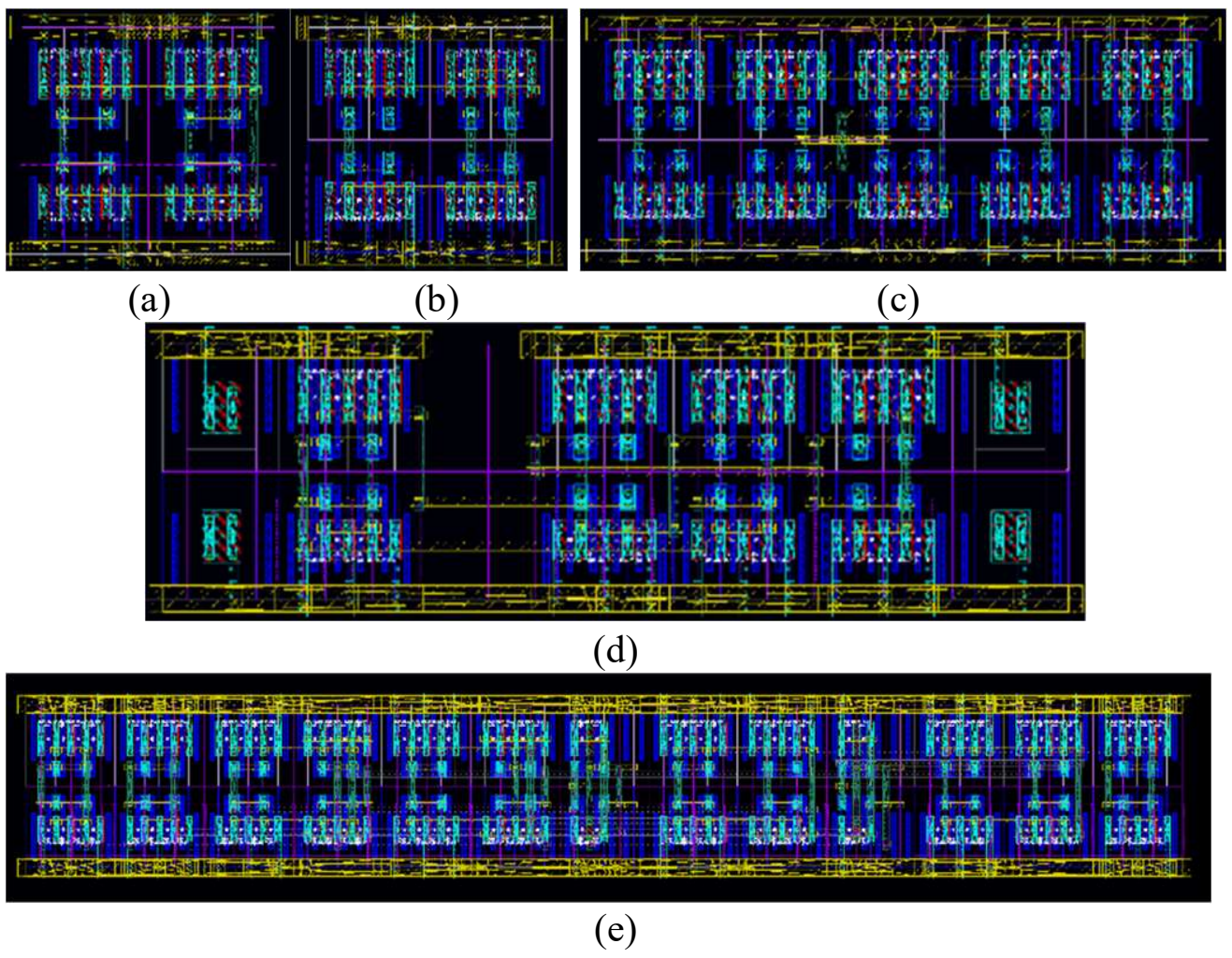

Fig. 6. Basic layout generation examples: (a) NAND (b) NOR (c) MUX (d) Level-Shifter (e) D Flip-Flop with Reset.

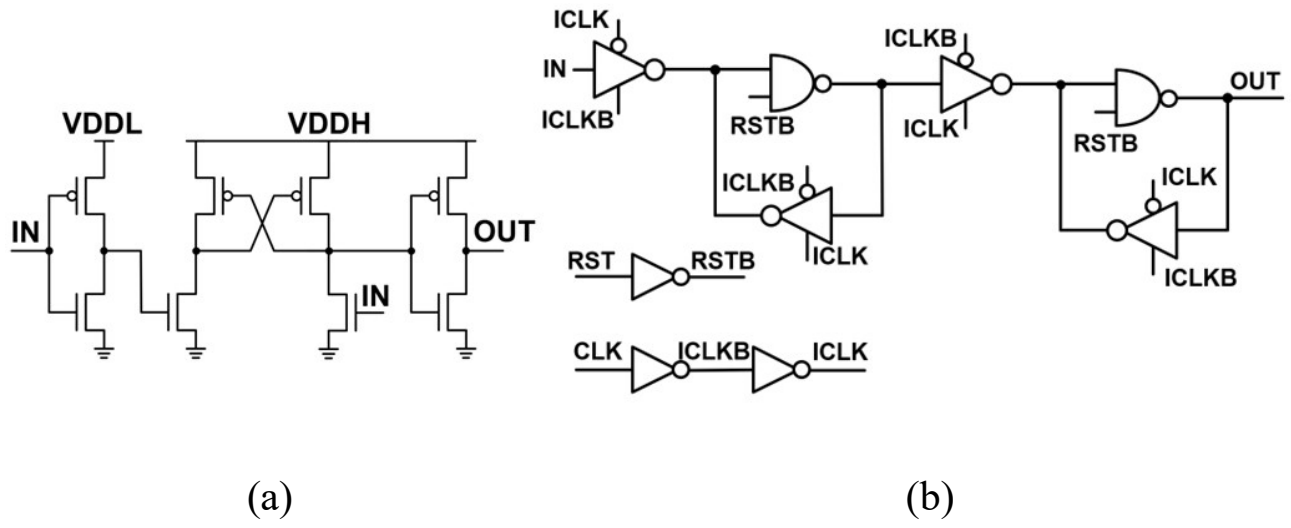

Fig. 7. Basic schematic examples: (a) Level Shifter (b) D Flip-Flop with Reset.

## IV. Layout Generation Results

The proposed interactive and automated layout generation process has been evaluated using a variety of design examples, from basic circuits like CMOS logic gates to advanced, high-performance circuits such as high-speed serializers for 100-Gb/s transmitters, as detailed in the sections that follow. The prompt logs for these examples are available in [16] for reference. The generated layouts from the proposed flow are illustrated in Fig. 6 and 10.

### *A. Basic Generation Examples*

*1) CMOS Logic Gates*: First, the basic functionality of our proposed approach is validated with primitive CMOS logic gates. In these experiments, we focused on optimizing transistor placement using naive algorithms guided by user feedback, interactively modifying device placements, and generating customized routing patterns. Flowcharts detailing these placement and routing processes are shown in Fig. 3 and 4. Given that CMOS logic gates often share similar placement and routing styles, such as positioning PMOS devices above NMOS devices and aligning transistors connected to similar nets, common strategies are used as prompts shared across designs. The LLM then interactively modifies the generated code based on user's modification instructions.

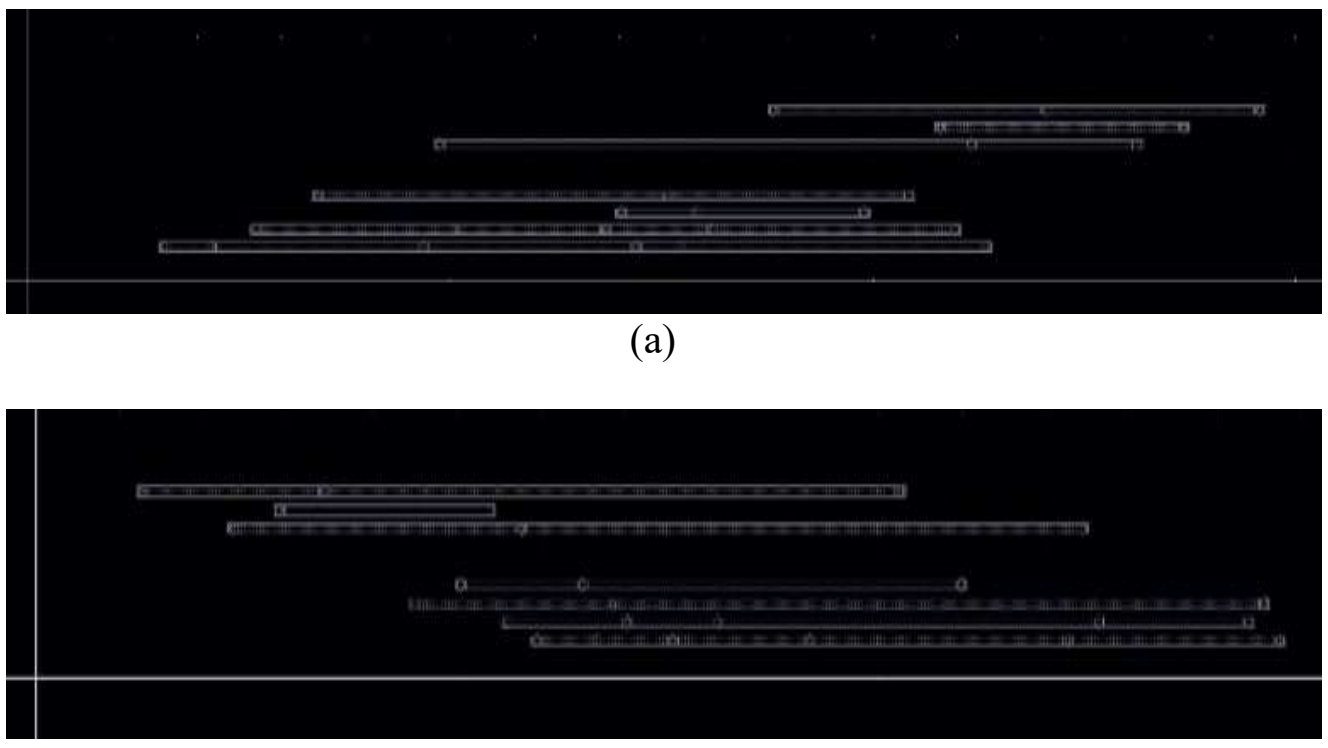

(a)

(b)

Fig. 8. Routing examples for a gate-level circuit: (a) when the instance order is optimized by LLM to meet user-defined objectives (b) when instances are placed in the order, they appeared in the input netlist.

*2) Complex Logic Gates*: Beyond simple gates, we have also generated more complicated circuits, such as multiplexers and level-shifters (Fig. 7(a)), which require more attention during routing between transistors. For smoother routing, additional customization in placement is necessary, which requires additional user-requested modifications. For example, in the multiplexer, due to 'twisted' routing between the EN signals, specific placement requests were made to accommodate these changes. The level-shifter underwent a similar process. If the initial prompt condition was not smoothly executed by ChatGPT, manual adjustments were made to the transistor arrangement. Subsequently, routing followed the process in Fig. 4, where, like the CMOS logic gates, it was often more effective for the designer to specify and modify the routing rather than relying solely on LLM automation. Here, a significant advantage was observed as LLMs could interactively apply modifications to the layout generator.

*3) Gate-Level Circuits*: Moving beyond transistor-level circuits, we also experimented with gate-level circuits, such as a CMOS D Flip-Flop with reset (Fig. 7(b)). It turns out that the gate-level circuits are simpler to process than transistor-level netlists because they are composed of gate instances placed in a row. LLMs are particularly effective in optimizing the order of the gate elements to satisfy user-defined constraints. Example constraints include placing instances with similar nets close to each other and positioning instances connected to input ports on the left (or right) side. Once the placement process is completed, routing wires are immediately generated by track-based routing functions supported by the layout generation framework. The track indices can be chosen by the LLM or users depending on customization requirements. As mentioned in the placement step, the LLM can understand the connectivity of instances and place the instances based on user-defined constraints, performance improvements can be achieved with a set of proper placement and routing instructions. For example, Fig. 8 shows the generated layout diagrams with and without the placement instruction to arrange instances with similar nets in proximity. It is observed that the LLM achieves approximately 18.6% reduction in routing wire length. This suggests the use of LLMs for simple optimization tasks without incurring significant time and cost expenditures. The layouts were created following the process depicted in Fig. 5. Users have the flexibility to adjust the placement of an instance or the location of a specific net, enabling customized placement and routing that optimizes layout performance without directly interacting with Application Programming Interfaces (APIs).

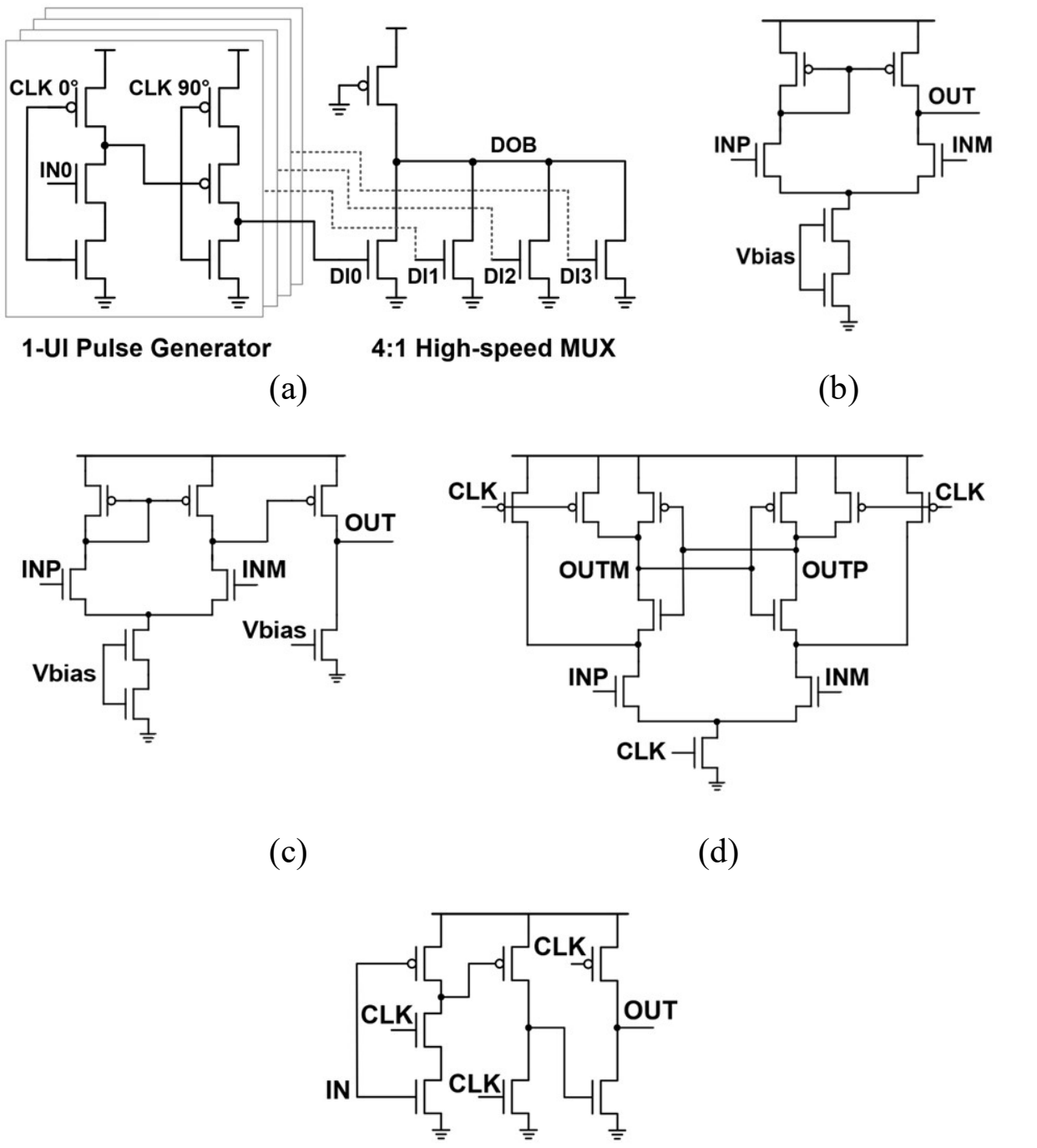

(a) (b)

(c) (d)

(e)

Fig. 9. Advanced circuit examples: (a) High-Speed 4-to-1 Serializer (b) Operational Transconductance Amplifier (c) Two-Stage Amplifier (d) Strong Arm Latch (e) Ratioed TSPC Flip-Flop.

### *B. Advanced Generation Examples - High-Speed Digital and AMS Circuits*

One of the principal advantages of the proposed approach is that designers articulate their intentions in natural language, which the LLM then translates into Python statements. These statements utilize functions defined in the layout generation API, enabling the construction of highly customized layout generators for high-performance applications. To demonstrate this capability, we generated an ultra-high-speed 4-to-1 serializer for wireline applications (Fig. 9(a)), an Operational

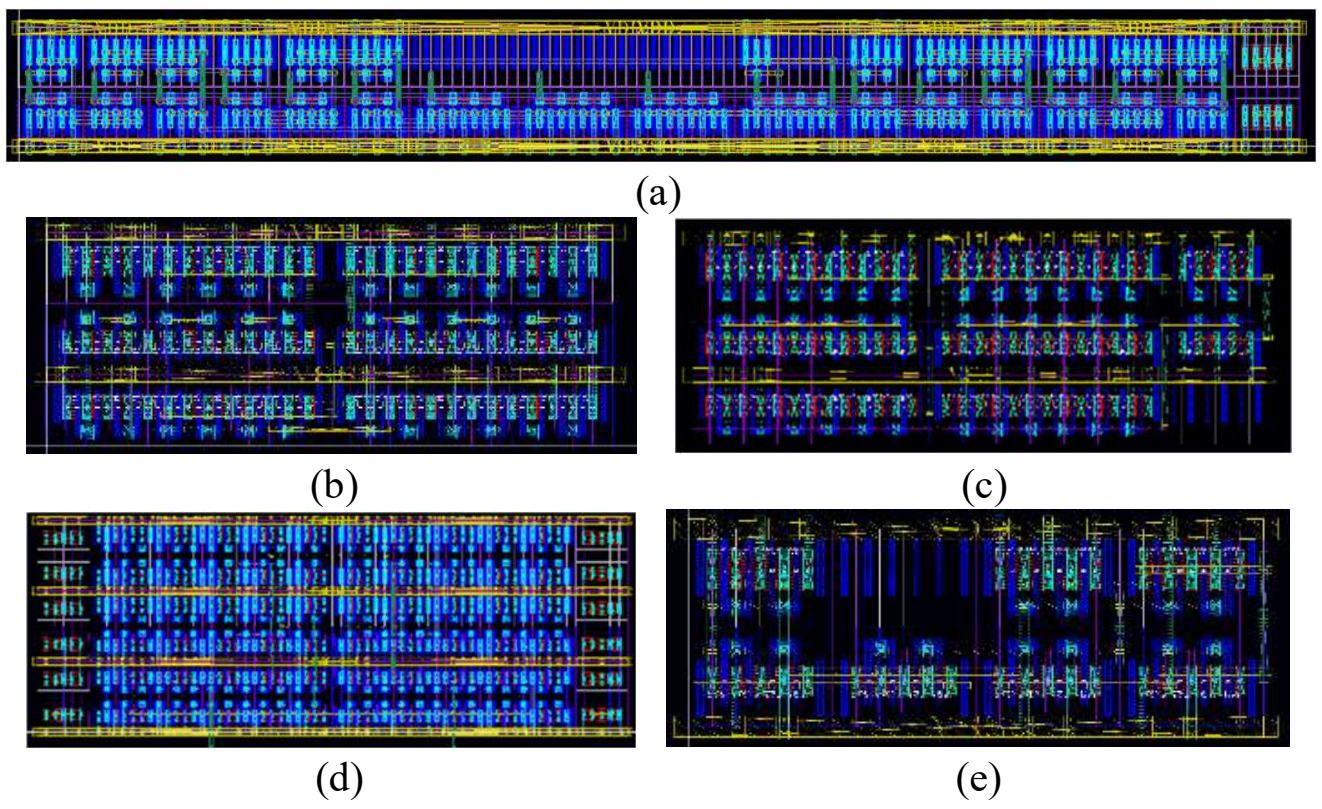

Fig. 10. Advanced layout generation examples: (a) High-Speed 4-to-1 Serializer (b) Operational Transconductance Amplifier (c) Two-Stage Amplifier (d) Strong Arm Latch (e) Ratioed TSPC Flip-Flop.

Transconductance Amplifier (OTA) commonly used in Low-Dropout Regulators (LDOs) (Fig. 9(b)), a two-stage amplifier (Fig. 9(c)) which based on OTA structure, a strong arm latch (Fig. 9(d)), a TSPC flip-flop which is a digital cell (Fig. 9(e)). While the generation of such highly customized circuits tends to require additional prompts for customization, our approach was still able to produce clean layout structures in terms of Design Rule Check (DRC) and Layout Versus Schematic (LVS) tests (Fig. 10).

## V. Evaluation and Future Work

The degree of optimization and interactive progress varies depending on the type of circuit (transistor vs. gate-level) and the complexity of operations (placement vs. routing). Typically, transistor-level circuits require a larger number of prompts due to their need for more detailed customization compared to gate-level circuits. It is also observed that the placement operations can perform basic optimizations, such as positioning instances with similar nets closed to each other, whereas routing operations for transistor-level circuits demand more interactive instructions due to their complexities and limited optimization capabilities, as summarized in Table I. Furthermore, to quantitatively compare the prompting efforts across various types of circuits and operations, the number of prompts present in each log was organized and analyzed. Although the number of prompts can also depend on the user's preference for organizing instructions, a large number of prompts does not necessarily mean a higher difficulty. Nevertheless, it is generally observed that the number of prompts tends to increase for designs with a larger number of instances or more complex routed nets, serving as an indicator of the prompt process (Table II).

The integrity of layout designs created from commands generated by LLMs was evaluated through Design Rule Check (DRC) and Layout Versus Schematic (LVS) tests, and it is observed that all generated layout designs passed these physical verification tests. Furthermore, we assessed the post-layout electrical performance of the high-speed 4-to-1 serializer in Fig. 11. The generated design successfully produced an output data stream at 32 Gb/s, matching the data rate reported in [17] using traditional design methods, and the simulation results are depicted in Fig. 11. This outcome reveals the potential of applying LLM-based layout automation methodologies to high-end applications, suggesting significant promise for future advancements in circuit design.

Overall, through this evaluation process, we confirmed the reliability and completeness of the layouts produced by LLMs, demonstrating the effectiveness of the proposed method. Additionally, the capability of interactive layout generation through user's intermediate commands provides a significant advantage for custom layout generation, enabling performance enhancements designated by the user. This is facilitated by the LLM's high correction ability and flexibility. However, it appears that the capacity of LLMs to autonomously create optimized layouts without user guidance, particularly in routing tasks for complex cells, requires further enhancement.

TABLE I. Qualitative Evaluation of Prompting for Various Layout Tasks

| Objective | Placement | Routing |
|---|---|---|
| User Interaction | Good | Good |
| Optimization | Normal | Bad |

TABLE II. Number of Prompts for Various Designs and Tasks

| Design | # of Instance | # of Nets | Operation | # of Prompts |
|---|---|---|---|---|
| NAND | 4 | 4 | Placement | 3 |
| | | | Routing | 3 |
| NOR | 4 | 4 | Placement | 3 |
| | | | Routing | 2 |
| Level Shifter | 8 | 5 | Placement | 3 |
| | | | Routing | 9 |
| 2-to-1 MUX | 10 | 11 | Placement | 8 |
| | | | Routing | 15 |
| D-FF with RESET | 9 | 7 | Placement | 7 |
| | | | Routing | 4 |
| D-FF with SET/RESET | 12 | 10 | Placement | 7 |
| | | | Routing | 4 |
| 4-to-1 SER Pulse Gen | 6 | 7 | Placement | 5 |
| | | | Routing | 12 |
| 4-to-1 SER MUX | 5 | 5 | Placement | 3 |
| | | | Routing | 10 |
| 4-to-1 SER Top | 5 | 8 | Placement | 3 |
| | | | Routing | 3 |
| OTA | 6 | 5 | Placement | 4 |
| | | | Routing | 10 |
| Two-Stage Amp | 8 | 7 | Placement | 4 |
| | | | Routing | 12 |
| Strong Arm Latch | 12 | 31 | Placement | 9 |
| | | | Routing | 36 |
| Ratioed TSPC-FF | 7 | 6 | Placement | 3 |
| | | | Routing | 11 |

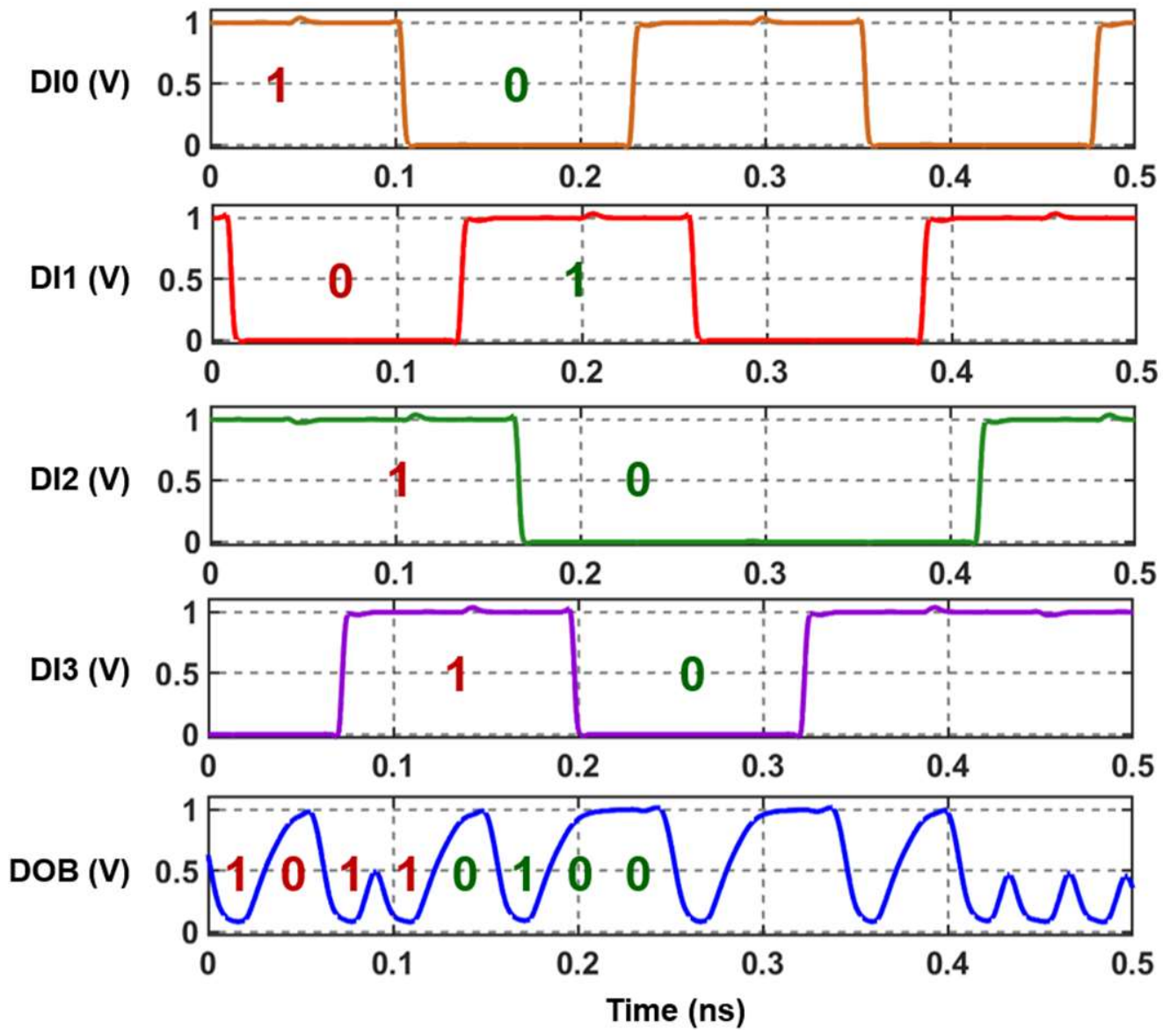

Fig. 11. Post-layout simulated waveforms of a 32Gb/s 4-to-1 Serializer.

Therefore, further research should be conducted on enhancing the self-optimization capabilities of LLMs for more complex tasks. In particular, it is essential to investigate suitable routing algorithms that can be integrated with LLMs to achieve optimization based on numerical metrics, such as total routed wire length. Additionally, there is a need for research into methods that seamlessly transition from placement to routing in a single process, simplifying the layout creation for users. This integration will enable users to more easily create optimal layouts using LLMs, offering substantial benefits in circuit design.

## VI. Conclusion

This study successfully demonstrates the application of Large Language Models (LLMs) for the automated generation of custom circuit layouts via an intuitive and natural-language-based interface. Our approach leverages a template-and-grid-based framework to create process-portable layout generators, adaptable to various custom circuits including standard cells and high-speed mixed-signal circuits. By utilizing LLMs, we significantly enhance the accessibility and efficiency of the layout generation process, moving away from the complexities typically associated with traditional programming languages. Additionally, this method allows for interactive and dynamic modifications of the layout generator code, thus improving customization capabilities to meet specific design requirements. Our experimental evaluations confirm that the LLM-based method can efficiently produce a diverse range of circuit layouts, offering extensive customization possibilities while ensuring robust DRC/LVS coverages. This work opens new avenues for further research into integrating LLMs in electronic design automation, promising to simplify complex design processes and boost productivity in circuit layout generation.